\documentstyle[preprint,aps]{revtex}

\input epsf

\tighten

\begin{document}

\draft

\title{NMR quantum computation with indirectly coupled gates}
\author{ David~Collins,$\mbox{}^{1}$ K.~W.~Kim,$\mbox{}^{1,}$\cite{kiwook} W.~C.~Holton,$\mbox{}^{1}$ H. Sierzputowska-Gracz,$\mbox{}^{2}$ and E. O. Stejskal$\mbox{}^{3}$ }
\address{ $\mbox{}^{1}$ Department of Electrical and Computer Engineering,
          Box 7911,
        North Carolina State University, Raleigh, 
        North Carolina 27695-7911 \\
         $\mbox{}^{2}$Department of Biochemistry,
          Box 7622,
          North Carolina State University, Raleigh, 
        North Carolina 27695-7622 \\
        $\mbox{}^{3}$Department of Chemistry,
          Box 8204,
          North Carolina State University, Raleigh, 
        North Carolina 27695-8204}

\date{\today}
      
\maketitle    

\begin{abstract}
An NMR realization of a two-qubit quantum gate which processes quantum information indirectly via couplings to a spectator qubit is presented in the context of the Deutsch-Jozsa algorithm. This enables a successful comprehensive NMR implementation of the Deutsch-Jozsa algorithm for functions with three argument bits and demonstrates a technique essential for multi-qubit quantum computation.
\end{abstract}



Nuclear magnetic resonance  (NMR) spectroscopy has emerged at the forefront of experimental quantum computation investigations \cite{chuang,jones,linden,chuang2,cory1,cory2,knill,cory3,madi,marx}. Key concepts such as fundamental gates \cite{chuang,cory1,madi} and error correction \cite{knill} have been demonstrated using NMR spectroscopy. However, comprehensive algorithm realizations have only been accomplished for the Deutsch-Jozsa algorithm for functions with one and two bit arguments \cite{chuang,jones,linden} and for Grover's algorithm with a two bit register \cite{chuang3,yannoni,jones2}. In these instances crucial two-qubit gates were realized ``directly'' via the coupling between the corresponding nuclear spins. For a quantum computer with a larger numbers of qubits, the associated requirement of appreciable coupling between any pair of spins raises difficulties. First, it may not be possible to find any molecule with this coupling configuration. Second, it demands increasingly complex schemes for managing the evolution of spectator spins during execution of any two-qubit gate \cite{lindenkupce}. 
However, efficient ``indirect'' realization of any two-qubit gate via a chain of couplings through intermediate spins is possible \cite{lloyd}. This relaxes the coupling requirements and a ``linear'' coupling configuration, for which the pattern of couplings is $\; A - B - C - D - ...$ suffices for quantum computation.  
In this Letter we report a comprehensive three qubit NMR realization of the Deutsch-Jozsa algorithm for functions with three bit arguments, using indirect realizations of two-qubit gates and a linear coupling configuration for information processing. The method that we present is general and readily scalable to larger numbers of qubits.

The Deutsch problem \cite{deutsch,cleve} considers $f:\{0,1\}^N \rightarrow \{0,1\}$ that are constant or balanced.  A {\em balanced} function returns $0$ as many times as $1$ after evaluation over its entire range. Given any $f$ which is either balanced or constant the problem is to determine its type. For classical algorithms that solve the Deutsch problem with certainty the number of evaluations of $f$ grows exponentially with $N$.
A quantum algorithm requires a single evaluation of $f$ and yet solves the problem with certainty \cite{deutsch,cleve}. Previous NMR demonstrations \cite{chuang,jones,linden} used the Cleve version \cite{cleve}, requiring an $N$ qubit control register for storing function arguments,  plus a $1$ qubit function register for function evaluation. Our recent modification to the algorithm (see Fig. \ref{pic:djscheme}) only needs the $N$ qubit control register \cite{collins}. It was also shown  that for each admissible (i.\ e.\ constant or balanced) function for $N \leq 2$ the evolution step is a product of single qubit operations. For an isolated spin $\frac{1}{2}$ nucleus any single qubit operation amounts to a rotation of the magnetization vector; this can be implemented classically.  Thus for $N \leq 2$ the algorithm can be executed classically. Previous comprehensive NMR implementations of the algorithm fall within this classical regime \cite{chuang,jones,linden}. However, for $N \geq 3$ there exist balanced functions for which qubits in the control register become entangled; this is indisputably quantum mechanical. $N=3$ is then the critical point at which quantum mechanical features become essential in the Deutsch-Jozsa algorithm. 

Two-qubit, entangling operations only appear during the function evaluation step, $\hat{U}_f$. However, for $f$ constant, $\hat{U}_f = \hat{I}$ and the algorithm can be executed classically. To assess quantum behaviour the corresponding gate for balanced $f$ must be investigated. Balanced functions can be characterized by the choices of $2^{N-1}$ arguments out of $2^N$ possibilities for which $0$ is returned. For $N=3$ this gives ${8 \choose 4} =  70$ balanced functions. Admissible functions may be represented via power series expansions in the argument bits, $x_i$ where $i=0,1,2$. That $x_i^2 = x_i$ for $x_i \in \{ 0,1 \}$ implies that for any admissible function,
\begin{equation}
  f(x_2,x_1,x_0) = \sum_{i>j \geq 0}^2 a_{ij} x_i x_j \oplus \sum_{i=0}^2 a_i x_i \oplus a
  \label{eq:seriesdecomp}
\end{equation}
where addition is modulo 2 and $a_{ij},a_i,a \in \{ 0,1 \}$. The disappearance of a cubic term in Eq. (\ref{eq:seriesdecomp}) is a property of balanced and constant functions. This provides a preliminary decomposition: 
\begin{equation}
 \hat{U}_f = \prod_{i>j \geq 0}^2 \left( \hat{U}^{ij} \right)^{a_{ij}}
             \prod_{k=0} \left( \hat{U}^{k} \right)^{a_{k}}
 \label{eq:fdecomp}
\end{equation}
where
\begin{eqnarray}
 \hat{U}^{ij}\left| x \right> & := & 
                         \left( -1\right)^{x_ix_j} \left| x \right>\mbox{, and} \\
 \hat{U}^i\left| x \right> & :=  & 
                         \left( -1\right)^{x_i} \left| x \right>.
 \label{eq:lingate}
\end{eqnarray}
are the {\em quadratic term gate} and {\em linear term gate}, respectively. The constant term merely provides an identity operation. Quadratic and linear term gates all commute and can be rearranged at will.
In terms of fundamental gates, $\hat{U}^i = \hat{R}^i_{\hat{\bf z}}(180)$ where superscripts index the qubit to which the $180^\circ$ single qubit rotation is applied. Similarly, $\hat{U}^{ij} =  \hat{R}^j_{-\hat{\bf y}}(90)
                                  \hat{U}^{ij}_{\mbox{\tiny CN}}
                                  \hat{R}^j_{\hat{\bf y}}(90)$
where $\hat{U}^{ij}_{\mbox{\tiny CN}}$ is a controlled-NOT gate with control $i$ and target $j$. Thus the algorithm can be implemented classically for functions with no quadratic terms. However, a quadratic term gate can produce  entangled states from unentangled states for qubits $i$ and $j$ via its constituent controlled-NOT gate. There are no classical operations involving magnetization vectors of two spins that produce entanglement. Therefore, for $N=3$ it is in the quadratic term gates that quantum mechanical features appear.
The arrangements of linear and quadratic term gates that constitute $\hat{U}_f$ for any admissible $f$ can be classified via similarity under permutations of control register qubits. Equations [(\ref{eq:fdecomp})-(\ref{eq:lingate})] imply that this can be accomplished by classifying admissible functions via similarity under argument bit permutations and/or addition of a constant term. Accordingly there are ten classes of balanced functions; a representative of each is provided in Table \ref{tab:foutput}. Function evaluation steps for any members of a given class differ only by a permutation of the control register qubits to which they are applied.
 Thus a  realization of the $N=3$ Deutsch algorithm is  comprehensive if the algorithm is applied to at least one representative from each class of admissible functions. All possible quadratic term gates are required for the classes represented by $f_9$ and $f_{10}$. Therefore, for $N=3$ the algorithm requires {\em two-qubit gates between all pairs of qubits} and is suitable for testing quantum information processing with a linear coupling configuration. 

Currently the most accessible experimental technology for quantum computing is NMR spectroscopy of spin $\frac{1}{2}$ nuclei of appropriate molecules in solution \cite{chuang,jones,linden,chuang2,cory1,cory2,knill,cory3,madi}. Any molecule containing three distinguishable, coupled spin $\frac{1}{2}$ nuclei in an external magnetic field provides the three qubits needed to solve the $N=3$ Deutsch problem.  To a good approximation the Hamiltonian for a room temperature solution state sample is $\hat{H} = \sum_{i=0}^2 \frac{\omega_i}{2} \hat{\sigma}^i_z 
                           + \frac{\pi}{2} \sum_{i>j \geq 0}^2 J_{ij}
                                  \hat{\sigma}^i_z \hat{\sigma}^j_z$, 
where $\omega_i$ are the Zeeman frequencies, $J_{ij}$ the scalar coupling constants, and $\hat{\sigma}^i_z$ Pauli operators \cite{slichter}. Superscripts label the spins and identify them with the corresponding argument bits. 
A literal translation of the algorithm into NMR operations would begin with initialization via a pseudo-pure state preparation scheme \cite{chuang2,cory1,cory2}. 
The evolution stage can be implemented by building single qubit and controlled-NOT gates from standard sequences of spin selective RF pulses and periods of evolution under $\hat{H}$ \cite{chuang2,cory1}.  Measurement can effectively be accomplished via tomography, which requires repeated execution of the initialization and evolution stages, each time varying the readout pulse before acquisition \cite{chuang2}.  It is, however, possible to apply the evolution stage directly to a thermal equilibrium initial state and successfully solve the Deutsch problem with an expectation value measurement \cite{linden,zhou}.  
Analysis of the system state after the function evaluation step demonstrates this for the $N=3$ Deutsch algorithm. Using the product operator formalism, the deviation part of the thermal equilibrium equilibrium density operator for a weakly coupled homonuclear NMR system is  proportional to $\hat{\rho}_{th} = \hat{I}^2_z + \hat{I}^1_z + \hat{I}^0_z$ \cite{slichter}. The initial rotation with phase $\phi = -\frac{\pi}{2}$ transforms this to $-\hat{I}^2_x - \hat{I}^1_x - \hat{I}^0_x$. Then $\hat{U}_f$ produces the states listed in Table \ref{tab:foutput}. Signal acquisition {\em immediately after application of $\hat{U}_f$} and with {\em no additional readout pulses} provides a spectrum, the {\em $f$-spectrum}, whose form depends on $f$. A {\em fiducial spectrum} is obtained in the same fashion but with $\hat{U}_f$ replaced by $\hat{I}$; here the system's pre-acquisition state is $-\hat{I}^2_x - \hat{I}^1_x - \hat{I}^0_x$. Comparison with $\hat{\rho}_f$ for admissible functions (see Table \ref{tab:foutput}) indicates that for each there is either a $0$ or a $\pi$ phase difference between each line each line of the $f$-spectrum and its counterpart in the fiducial spectrum. More precisely for applicable functions: 

(i) {\em $f$ is constant if and only if  the $f$-spectrum is identical to the fiducial spectrum and} 

(ii) {\em $f$ is balanced if and only if there is a $\pi$ phase difference between at least one line of the $f$-spectrum and its counterpart in the fiducial spectrum.} 

This criterion requires that each spin is coupled to at least one other spin. If any spin is completely uncoupled then the entire $f$-spectra for $f_9$ and $f_{10}$ disappears; the comparison would be impossible. However, if each spin is coupled to at least one other then for $N=3$ at least one is coupled to the other two. This ensures that at least one of the doubly antiphase multiplets for $f_9$ and $f_{10}$ survives, giving a line in the $f$-spectrum with a $\pi$ phase difference relative to its fiducial spectrum counterpart.  For $f_7$ and $f_8$ at least one of the antiphase multiplets (for spin 2 or spin 0) must survive; again this provides a line whose phase differs by $\pi$. For $f_4$, $f_5$ and $f_6$ the entire spin 0 multiplet in the $f$-spectrum displays a $\pi$ phase difference relative to its fiducial spectrum counterpart. The same is true for spin 2 in the cases $f_1$, $f_2$ and $f_3$.

The fiducial spectrum can be phased so that its constituent lines all appear upright. Thus, the answer to the $N=3$ Deutsch problem may be determined by inspecting the $f$-spectrum for inverted lines. Each balanced function produces at least one inversion. For constant functions all lines are upright.   This provides a solution state NMR scheme for conclusively answering the $N=3$ Deutsch problem with just one application of the evolution stage (here equivalent to the unmodified version followed by a $\hat{R}_{\bf \hat{n}}(90)$ readout) to the thermal equilibrium input state.
 
A saturated solution of $^{13}\mbox{C}$ labeled alanine in $\mbox{D}_2\mbox{O}$ provided the qubits. We label the carboxyl carbon, spin $2$, the $\alpha$ carbon, 1 and the methyl carbon, $0$. Protons were decoupled using a standard heteronuclear decoupling technique. Scalar couplings are $J_{21} = 56$Hz, $J_{10} = 36$Hz and  $J_{20} = 1.3$Hz. Relaxation times are $T_1(2)=11.5$s, $T_1(1)=1.2$s, and $T_1(0)=0.7$s and $T_2(2)=1.3$s, $T_2(1)=0.41$s, and $T_2(0)=0.81$s where the argument labels the spin. Spin selective rotations were implemented via Gaussian shaped pulses of duration $0.7$ms for spins 0 and 1 and $0.5$ms for spin 2. No hard pulses were used. Linear term gates can be implemented via spin selective phase shifts on the output spectrum by placing them after the quadratic term gates and immediately prior to acquisition. Thus certain $f$-spectra differ by spin selective phase shifts only. These are: (i) $f_1$, $f_2$ and $f_3$, (ii) $f_4$, $f_5$, and $f_6$, (iii) $f_7$ and $f_8$ and (iv) $f_9$ and $f_{10}$.  The crux of the experiment is in the realization of the quadratic term gates. This is accomplished via the rotation and delay construction of controlled-NOT gates \cite{chuang,jones,cory2}. Selective refocusing sequences \cite{linden2,leung2,jones3} effectively eliminate all but one coupling term in $\hat{H}$, thus providing appropriate evolution during the delay. The resulting quadratic term gate simplifies to
\begin{equation}
 \hat{U}^{ij} =   \left[ 1/2J_{ij} \right]^{ij}
                  - \left[ 90 \right]^i_{-z}
                  - \left[ 90 \right]^j_{-z}
\end{equation}
where $\left[ \theta \right]^j_n$ indicates a rotation of spin $i$ about the axis $n$ through angle $\theta$ and $\left[ t \right]^{ij}$ evolution under the scalar coupling between spins $i$ and $j$ for period $t$. For alanine this is satisfactory for $\hat{U}^{21}$ and $\hat{U}^{10}$. However, for $\hat{U}^{20}$ it is inadequate since $1/2J_{20} = 0.42$s which is comparable to the smallest $T_2$. An alternative is to process the information via spin 1 and use only the linear coupling configuration ( spin 2 - spin 1 -spin 0). An indirect realization is  $\hat{U}^{20}_{\mbox{\tiny CN}} =    \hat{U}^{01}_{\mbox{\tiny SW}}
           \hat{U}^{21}_{\mbox{\tiny CN}}
           \hat{U}^{01}_{\mbox{\tiny SW}},$ 
where $\hat{U}^{01}_{\mbox{\tiny SW}}$ is the SWAP gate \cite{madi} between qubits $1$ and $0$. After simplification,

\begin{eqnarray}
 \hat{U}^{20} & = &   \left[ 90 \right]^1_y
                 - \left[  90\right]^0_y
                 - \left[ 1/2J_{10} \right]^{10}
                 - \left[ 90 \right]^1_x 
                 - \left[ 90 \right]^0_x \nonumber \\
             &  &
                 - \left[ 1/2J_{10} \right]^{10}
                 - \left[ 90 \right]^1_y 
                 - \left[ 1/2J_{21} \right]^{21} 
                 - \left[ 90 \right]^1_x \nonumber \\
             &  &
                 - \left[ 1/2J_{10} \right]^{10}
                 - \left[ 90 \right]^1_y
                 - \left[ 90 \right]^0_x 
                 - \left[ 1/2J_{10} \right]^{10} \nonumber \\
             &  &
                 - \left[ 90 \right]^1_x 
                 - \left[ 90 \right]^0_{-y}
                 - \left[ 90 \right]^2_{-z}
                 - \left[ 90 \right]^3_{-z}.
\end{eqnarray}
This gives a pulse sequence of duration  $0.071s$ (excluding $\hat{\bf z}$ rotations that are equivalent to phase shifts in the output spectrum) that is  significantly faster than that using $\left[ 1/2J_{20} \right]^{20}$. During $\left[ 1/2J_{10} \right]^{10}$ and $\left[ 1/2J_{21} \right]^{21}$ evolution periods the selective refocusing scheme effectively removes the scalar coupling between spins 2 and 0. Throughout this implementation of the algorithm {\em information is processed using only the spin2 - spin 1 -spin 0 linear coupling configuration and not the spin 2 -spin 0 coupling}. However, the latter must be taken into account in the interpretation of the output spectra. 

The experiments were performed at room temperature using a BRUKER 500-DRX spectrometer and an inverse detection probe.  For each representative function listed in Table \ref{tab:foutput} signal acquisition takes place immediately after implementation of $\hat{U}_f$. Figure \ref{pic:output} provides selected experimental spectra that are phased so that $-\hat{I}_x^j$ product operator terms correspond to upright multiplets. The line orientations agree with those predicted from $\hat{\rho}_f$ and provide correct solutions to the Deutsch problem. 
An estimate of errors for the most complicated case, $f_9$, may be obtained by applying a selective $90^\circ$ readout pulse about the $\hat{\bf x}$ axis immediately after $\hat{U}_f$. Ideally the readout spin multiplet should remain while the others disappear.  The average amplitudes of the residual signals for the latter lie between $14\%$ and $31\%$ of the average amplitude of the corresponding lines with no readout. 
The ability to extract the Deutsch problem solution for $N > 3$ via pure phase information in the output spectrum, in contrast to  amplitude information, appears to have mitigated such errors. It is not yet clear how this advantage may be extended beyond $N = 3$.
 The number of selective rotations required for $f_9$ points to imperfections within selective pulses as one likely source of error. In particular, it must be noted that possible effects of scalar coupling evolution during application of selective rotations were ignored. Indeed, for the indirectly coupled realization of $\hat{U}^{20}$ the total duration of all the selective rotations is comparable to $1/2J_{21}$. To the best of our knowledge this issue has not been addressed satisfactorily. A further possible source of error are inhomogeneities in the RF magnetic fields used for selective rotations. 

To conclude, we have provided a comprehensive NMR realization of the $N=3$ Deutsch-Jozsa algorithm, at the same time demonstrating quantum information processing via a linear configuration of couplings and indirect realizations of two-qubit gates. The use of appropriate SWAP gates allows for the extension of our method to quantum computation with larger numbers of qubits.

This work was supported, in part, by the DARPA and the ONR. We would also like to thank Gary Sanders for useful discussion.

\begin{figure}[h]
                  \caption{Deutsch-Jozsa algorithm. Initialization to the state:  
$\left| x \right> \equiv \left| x_{N-1} \right> ... \left| x_0 \right>$
where $x_i \in \{ 0,1 \}$.
$\hat{R}_{{\bf n}} (90)$ rotates each qubit through $90^\circ$  about  ${\bf n} = \cos{\phi}\hat{\bf x} + \sin{\phi}\hat{\bf y}$. For a single qubit
  $\hat{R}_{{\bf n}}(\theta) := e^{-i{\bf n}.\hat{\bf \sigma} \theta/2}$
 where $\hat{\bf n}.\hat{\bf \sigma} = n_x \hat{\sigma}_x + n_y \hat{\sigma}_y + n_z \hat{\sigma}_z$. The $f$-controlled gate, $\hat{U}_f \left| x \right> := \left( -1\right)^{f(x)} \left| x \right>$, evaluates $f$. The expectation value of $\left| x \right>\left< x \right|$ on the output state answers the problem conclusively. }  
                  \label{pic:djscheme}
\end{figure} 

\begin{figure}
  \caption{$\mbox{}^{13}$C output spectra for alanine: (a) Fiducial spectrum, (b) $f_2$ output spectrum, (c)$f_4$ output spectrum, (d) $f_7$ output spectrum, and (e) $f_9$ output spectrum. Insets provide enlargements of all antiphase and doubly antiphase multiplets.}
  \label{pic:output}
\end{figure}

\begin{table}
  \caption{Representatives of each class of admissible functions ($N=3$) and the corresponding density operator after the function evaluation step, $\hat{\rho}_f := \hat{U}_f \hat{R}_{\bf -\hat{y}}(90) \hat{\rho}_{th}$.
The form of the spin $j$ multiplet depends on the term containing a factor of $\hat{I}^j_x$. Line intensities within a given multiplet are equal. $\hat{I}^j_x$ terms yield a multiplet with lines of equal phase. $\hat{I}^i_z \hat{I}^j_x$ yields an antiphase multiplet if $J_{ij} \neq 0$.  If $J_{ij}=0$ antiphase lines overlap and the entire multiplet disappears. $\hat{I}^i_z \hat{I}^j_x \hat{I}^k_z$ yields a doubly antiphase multiplet for spin $j$ if $J_{ij},J_{jk} \neq 0$. If $J_{ij} = 0$ or $J_{jk} = 0$ the multiplet again disappears. Antiphase and doubly antiphase multiplets contain lines whose phases differ by $\pi$. Multiplication by $-1$ corresponds to a phase shift of $\pi$ for each line in the corresponding multiplet.} 
 \begin{tabular}{ll}
  Representative, $f$   & $\hat{\rho}_f$  \\ \hline
  Constant: &            \\
  $f_{const} = 0$   & $-\hat{I}^2_x - \hat{I}^1_x - \hat{I}^0_x$ \\ 
  Balanced: &            \\
  $f_1 = x_2$                           & $+\hat{I}^2_x - \hat{I}^1_x - \hat{I}^0_x$ \\
  $f_2 = x_2 \oplus x_1$                & $+\hat{I}^2_x + \hat{I}^1_x - \hat{I}^0_x$ \\
  $f_3 = x_2 \oplus x_1 \oplus x_0 $    & $+\hat{I}^2_x + \hat{I}^1_x + \hat{I}^0_x$ \\
  $f_4 = x_2 x_1 \oplus x_0$            & $-2 \hat{I}^2_x \hat{I}^1_z
                                           -2 \hat{I}^2_z \hat{I}^1_x 
                                           + \hat{I}^0_x$ \\
  $f_5 = x_2 x_1 \oplus x_2 \oplus x_0 $ & $+2 \hat{I}^2_x \hat{I}^1_z
                                          -2 \hat{I}^2_z \hat{I}^1_x 
                                           + \hat{I}^0_x$\\
  $f_6 = x_2 x_1 \oplus x_2 \oplus x_1 \oplus x_0 $ & $+2 \hat{I}^2_x \hat{I}^1_z
                                                       +2 \hat{I}^2_z \hat{I}^1_x 
                                                       + \hat{I}^0_x$\\
  $f_7 = x_2 x_1 \oplus x_1 x_0 \oplus x_2 \oplus x_1$ & $+2 \hat{I}^2_x \hat{I}^1_z
                                                        +4 \hat{I}^2_z \hat{I}^1_x \hat{I}^0_z 
                                                        -2 \hat{I}^1_z \hat{I}^0_x$\\
  $f_8 = x_2 x_1 \oplus x_1 x_0 \oplus x_2 $ & $+2 \hat{I}^2_x \hat{I}^1_z
                                                -4 \hat{I}^2_z \hat{I}^1_x \hat{I}^0_z 
                                                -2 \hat{I}^1_z \hat{I}^0_x$ \\
  $f_9 = x_2 x_1 \oplus x_1 x_0 \oplus x_2 x_0 $ & $-4 \hat{I}^2_x \hat{I}^1_z \hat{I}^0_z 
                                                    -4 \hat{I}^2_z \hat{I}^1_x \hat{I}^0_z 
                                                    -4 \hat{I}^2_z \hat{I}^1_z \hat{I}^0_x$ \\
  \hspace{-0.4em}$\begin{array}{cl}
     f_{10} =  & x_2 x_1 \oplus x_1 x_0 \oplus x_2 x_0  \\
               & \oplus x_1 \oplus x_0 
   \end{array}$
         &$-4 \hat{I}^2_x \hat{I}^1_z \hat{I}^0_z 
           +4 \hat{I}^2_z \hat{I}^1_x \hat{I}^0_z 
           +4 \hat{I}^2_z \hat{I}^1_z \hat{I}^0_x$ 
 \end{tabular}

\label{tab:foutput}
\end{table}

\pagebreak

\thispagestyle{empty}

\vspace{1in}

\begin{figure}[h] 
  \centerline{\epsffile{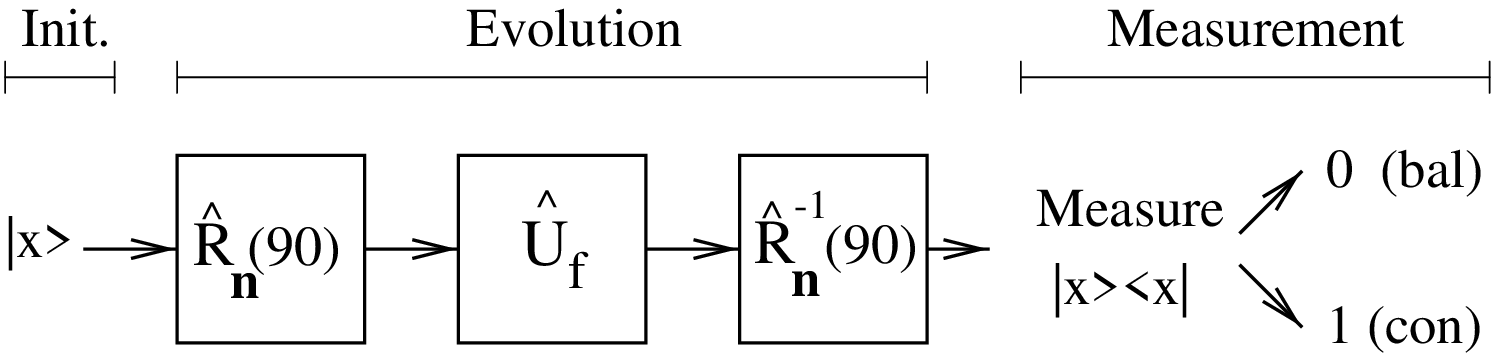}}
\end{figure} 

\vspace{4in}

\centering {\large Figure 1}

\pagebreak
\thispagestyle{empty}

\begin{figure}
  \epsfysize=6in
  \centerline{\epsffile{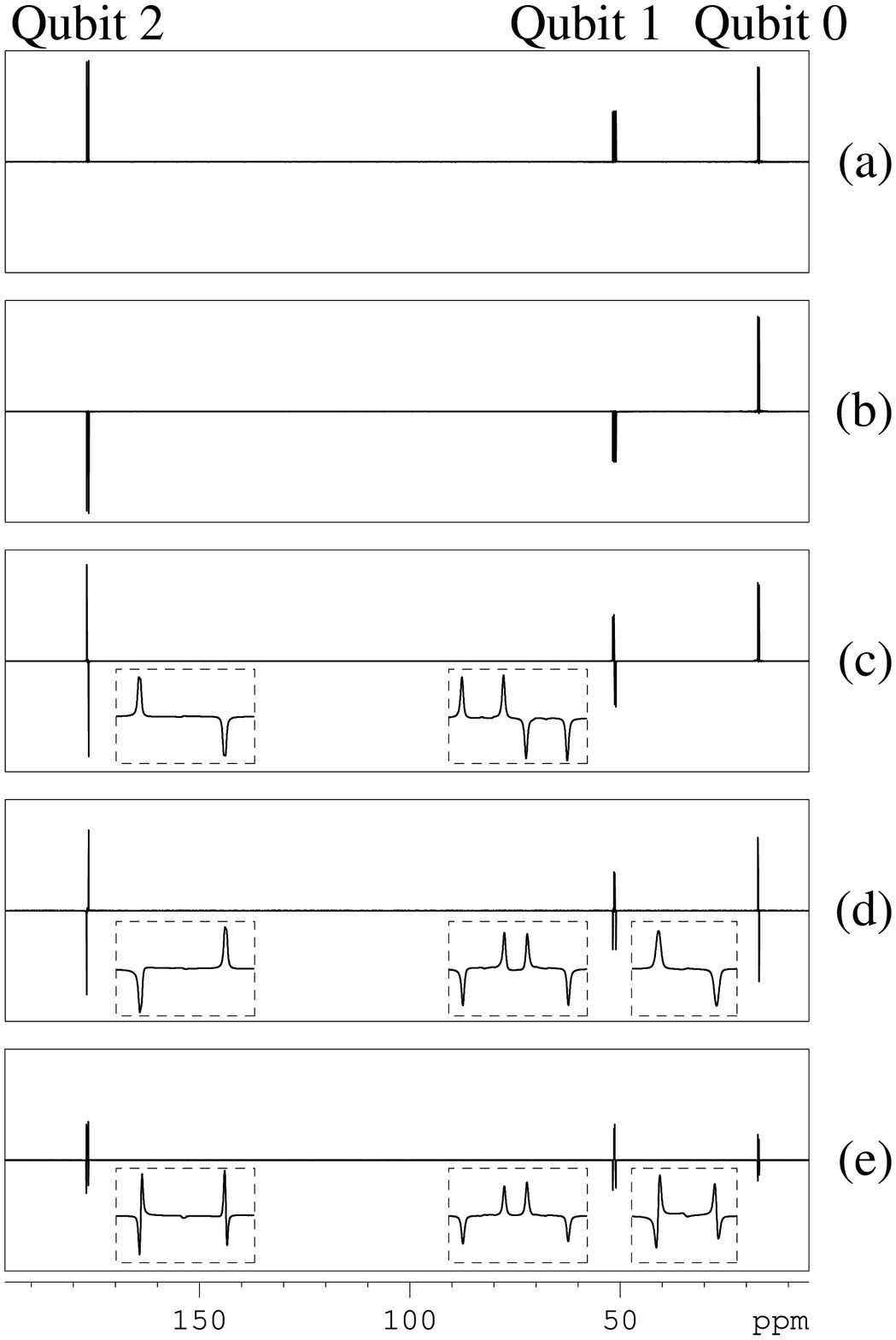}}
\end{figure}

\vspace{2in}

\centering {\large Figure 2}
 
\end{document}